\documentclass[
aip,
jmp,
reprint,
]{revtex4-1}

\pdfoutput=1

\usepackage{graphicx}
\usepackage{dcolumn}
\usepackage{bm}

\usepackage{float}
\usepackage[font={small}]{caption}

\usepackage{amsfonts,amssymb,amsmath}
\usepackage{subfigure}
\usepackage{tabularx}
\usepackage[font={footnotesize}]{caption}
\usepackage{framed}
\usepackage{algorithm}
\usepackage{algorithmicx}

\usepackage{enumitem}
\usepackage{calc}

\usepackage{changepage}

\usepackage{marginnote}

\usepackage{verbatim}

\usepackage{framed}
\usepackage{rotating}

\usepackage{color}

\graphicspath{{./images/}} 

\begin{document}

\preprint{AIP/123-QED}

\title{Finding metastable states in real-world time series with recurrence networks}
\author{I. Vega}
\email{iliusi@math.fu-berlin.de}
\affiliation{Mathematics Institute, FU Berlin, Gernany}

\author{Ch. Sch\"utte}
\affiliation{Mathematics Institute, FU Berlin, Gernany}

\author{T. O. F. Conrad}
\affiliation{Mathematics Institute, FU Berlin, Gernany}

\begin{abstract}
\small{In the framework of time series analysis with recurrence networks, we introduce a self-adaptive method that determines the elusive recurrence threshold and identifies metastable states in complex real-world time series. As initial step, we introduce a way to set the embedding parameters used to reconstruct the state space from the time series. We set them as the ones giving the maximum Shannon entropy for the first simultaneous minima of recurrence rate and Shannon entropy. To identify metastable states, as well as the transitions between them, we use a soft partitioning algorithm for module finding which is specifically developed for the case in which a system shows metastability. We illustrate our method with two complex time series examples. Finally, we show the robustness of our method for identifying metastable states. Our results suggest that our method is robust for identifying metastable states in complex time series, even when introducing considerable levels of noise and missing data points.}

\end{abstract}

\keywords{recurrence quantification analysis; metastability; non-linear dynamics; threshold}

\pacs{ 05.45.Tp, 05.40.-a, 05.45.-a, 07.05.Kf, 89.75.Fb, 89.75.Kd, 89.75.Hc}

\maketitle


\section{Introduction}\label{sec:Introduction}

The need to understand the dynamics of complex data coming from the biological, the financial, the environmental or the medical fields, has promoted the development of many visualization and analysis methods. According to~\cite{Maat09}, some of the linear methods$\,$--$\,$such as PCA or Classical Multi-dimensional Scaling$\,$--$\,$and non-linear methods$\,$--$\,$such as Stochastic Neighbor Embedding or Isomaps$\,$--$\,$used for this purpose, can have some drawbacks, like not preserving both local and global scale properties of complex data or depending on many undetermined parameters. These problems can leave large part of the analysis open to subjective interpretation.

An alternative approach that gives information about the local, medium and global scales in high-dimensional, non-linear time series, is recurrences analysis via recurrence plots and recurrence networks. A recurrence plot is a tool to visualize phase space trajectories which provides dynamical information of even high-dimensional data sets. A recurrence network is the graph representation of a recurrence plot. The theoretical foundations of these constructions are explained in Section \ref{sec:Background}.

One of the difficulties of computing a recurrence plot is selecting its recurrence threshold. The recurrence threshold is a parameter that controls how close two phase space trajectories, or state space vectors, should be in order to consider them as neighbors. Therefore, it determines the size of neighborhoods in phase space, that can ultimately be associated with the existence of metastable dynamical states.

How to set an appropriate recurrence threshold for real-world time series is a problem long discussed~\cite{Donn10, Schi11}. This problem originates in some common properties of real-world time series: having a non-necessarily uniform probability distribution, frequently having noise or missing some measurement points, and showing \textit{metastability} (see Section \ref{sec:Background}).

In this paper we introduce a self-adaptive method for the identification of metastable states in real-world time series. Our method is broadly divided into three parts, explained in Section \ref{sec:Method}. The first component is devoted to the estimation of the parameters necessary to construct the state space from a given time series. These parameters are set in terms of the simultaneous first minima of two recurrence quantitative analysis (RQA) measurements: Shannon entropy and recurrence rate. In the second part of our method, we compute an appropriate recurrence threshold for the construction of a recurrence network. The idea behind this computation is that a robust recurrence threshold should be located in a region of values that provide similar modular structure in their associated recurrence networks. Finally, we identify metastable states in the time series, as well as the transitions between them.

The performance of our method is illustrated in Section \ref{sec:Results}, where we apply it to two time series potentially showing metastability. In Section \ref{sec:Robustness} we validate its ability to identify metastable states in a robust way.

To the best of our knowledge, this is the first time the concept of metastability is introduced in the analysis of real-world time series with recurrence networks. The adaptive quality of out method enables the selection of embedding parameters and recurrence threshold in a robust way. The results of analyzing real-world time series with out method are also robust to the addition of noise and missing data points.


\section{\label{sec:Background}Background}

A state space trajectory is a series of states describing the evolution of a system. Since any dynamical system can be defined by its state space and an evolution operator, the time evolution of the state space trajectory explains the dynamics of the system.

In a dissipative dynamical system, even a small perturbation of the state at a given time, can cause an exponential divergence in a future state. However, for sufficiently long time, this system will return to a state arbitrarily close to a former state. This return is called a recurrence.

A recurrence plot is a tool to represent recurrences of state space trajectories $\mathbf{x}_i$ to the neighborhood of a set of states. The size of each neighborhood is given by the \emph{recurrence threshold}, $\varepsilon$. This way, a recurrence plot is defined as a square binary matrix, $R_{ij}(\varepsilon)$
\begin{equation}
R_{ij}(\varepsilon)=\Theta \left( \varepsilon - d(\mathbf{x}_i, \mathbf{x}_j) \right)-\delta_{ij} \label{eq:RP}
\end{equation}
Where $\Theta (\cdot)$ is a Heaviside function and $d(\mathbf{x}_i, \mathbf{x}_j)$ is a metric between two state space vectors $\mathbf{x}_i$ and $\mathbf{x}_j$.

The state space trajectory of a dynamical system underlying a time series data, can be reconstructed from the time series using embedding techniques. We will use the delay mapping method, based on Taken's theorem of embedding~\cite{Take81}. This method requires setting two parameters: the embedding delay, $\tau$, and the embedding dimension, $m$~\cite{Whit36,Abar96}. Different selections of embedding parameters will reconstruct state spaces with different dynamical information quality. For a time series $u_i$ of length $N$, using the delay mapping method, the $N^* = N - \tau(m-1)$ state space trajectories reconstructed, $\mathbf{x}_i$, are given by
\begin{equation}
\mathbf{x}_i = (u_i, u_{i+\tau}, ..., u_{i+(m-1) \tau}) \mbox{, for } i = 0, ..., N^*
\label{eq:phaseSpaceTrajectory}
\end{equation}

In order to set the embedding delay, one must guarantee that the vector built from all the $i$-th entries of the state space trajectories is linearly independent from the vector built from all $j$-th entries of the state space trajectories, for all $i \neq j$. This implies that for periodic time series, the embedding delay can not be a multiple of the period. Thus, the embedding delay can be set in terms of the linear autocorrelation function or of the average mutual information. For an extended discussion on how to determine the embedding time delay, see the work of Abarbanel~\cite{Abar96}.

To set the embedding dimension, different geometrical, dynamical and topological tests can be used~\cite{Lete08}. The geometrical tests, like the computation of fractal dimensions or false nearest neighbors, indicate the variations in distance between two close points when the embedding dimension increases. The dynamical tests, like the implementation of predictability tests or the estimation of Lyapunov exponents, are used to select the embedding that provides a unique future for every data point. The topological tests look for the embedding dimension $m$ that avoids intersections of stable periodic orbits. For more information on embeddings, see the articles of Adachi~\cite{Adaci93} and Sauer et al.~\cite{Saue91}.

The introduction of recurrence quantitative analysis (RQA) measurements by Zbilut and Webber in the early nineties~\cite{Zbil92, Webb94}, broadened the concept of recurrence. Marwan et al.~\cite{Mar02,Mar02b} studies of the geometrical interpretation of these measurements indicated that recurrence plots were a convenient tool to analyze non-linear data~\cite{Mar07}. They also opened the door to the potential analysis of high-dimensional~\cite{Marw13, Donn10b} and non-stationary~\cite{Chen12, DongesPhD, MarwanPhD} time series.

The RQA measurements give information about the local, medium and global scales of a dynamical system. Some of these are typically not invariant to changes on the embedding parameters, but dynamical invariants like correlation entropy and correlation dimension can also be derived from recurrence plots~\cite{Mar07}. 

In 2008, seemingly independently, Krishnan et al.~\cite{Kris08a,Kris08b}, Xu et al.~\cite{Xu08} and Yang and Yang~\cite{Yang08} introduced the concept of recurrence network. In a recurrence network, every node represents one of the state space vectors associated to the time series and every edge represents the belonging of a pair of state space vectors to a same recurrence neighborhood. Due to the formulation of recurrence plots, recurrence networks are unweighted, undirected and have the same number of nodes as the number of state space vectors built from the data set, $N^* = N - \tau(m-1)$.

Donner et al. have proved that the local, medium and global geometric properties of a system can also be recovered from the recurrence network through measurements based on neighborhoods or on paths~\cite{Donn11}.

Since the structure of a recurrence network depends on the closeness between state space vectors, recurrence regions should originate dense groups of interacting nodes in the network, or modules~\cite{Bruc13}. The connectivity of a recurrence network, meaning the size and number of modules it contains, can be modified by varying the recurrence threshold.

How to set an appropriate recurrence threshold when analyzing real-world time series is a problem long discussed~\cite{Donn10, Schi11}. This problem originates in some common properties of real-world time series: non-necessarily uniform probability distribution, noise or missing data points, and showing \textit{metastability}.

Metastability is a property of physical phenomena with multiple, well-separated time scales~\cite{Sari12}. At a short time scale, a dynamical system showing metastability seems to be in equilibrium. In this case, one can identify so-called metastable states. At a different, longer time scale, this system does not seem to be in equilibrium because it undergoes transitions between the metastable states. For an extended review on metastability, see the work of A. Bovier~\cite{Bovi09}.

Initially, the recurrence threshold was set ``using rules of thumb''~\cite{Donn10b,Donn11} over the variation of RQA measurements~\cite{Zbil02}, the diameter of the reconstructed state space, some dynamical measurements such as the correlation integrals~\cite{Zou10}, correlation dimensions~\cite{Grass83}, second order R\`enyi entropy~\cite{Thie04,Schu11} or attractor dimensions~\cite{Donn10}. Generally, the recurrence threshold was kept as small as possible.

Later on, graph theory concepts were introduced to the study of recurrence networks in order to address the problem of selecting the appropriate recurrence threshold. A summary of the problems associated to the selection of this parameter is given in~\cite{Donn10}.

Feldhoff et al.~\cite{Feld12} selected recurrence thresholds that produced recurrence networks with low edge densities, because higher edge density values tend to hide important dynamical structures. Additionally, they asked for values of the parameter such that a small variation in the recurrence threshold did not produce noticeable differences in the dynamical analysis results.

In 2012, Donges et al.~\cite{Dong12} introduced an analytical framework based on random geometric graphs~\cite{Dall02}. Considering RGG theory, they determined the recurrence threshold for one-dimensional time series with uniform probability density distribution in terms of the percolation threshold $\epsilon_c$. This threshold indicates the limit in which the network's giant component breaks down and makes impossible to recover information about mesoscopic and path-based measures~\cite{Pen03}. This boundary for the recurrence threshold considers that for too large $\varepsilon$, the recurrence network becomes too dense, and for too small $\varepsilon$ the recurrence network's giant component breaks down into smaller disconnected components. In both cases the fine geometry of the time series is not well represented by the neighborhood- and path-measurements. This way, Donges et al. focused on the study of the average path length, which relates to the network's giant component, to set a range of values for the recurrence threshold. However, there are no exact analytical results for $d$-dimensional random geometric graphs of arbitrary $d$ and the authors suggest returning to the results of Donner et al.~\cite{Donn10,Donn11} for general cases. 

Some approaches to the analysis of  recurrence networks constructed from time series with non-uniform distributions are the study of changes in connectivity by Hsing and Rootz\'e~\cite{Hsin05}, and more recently by Cooper and Frieze~\cite{Coop13}. On the other hand, Kong and Yeh~\cite{Kong07} have investigated the problem of characterizing the critical density and critical mean degree of random geometric graphs with non-uniform probability distributions. Based on probabilistic methods and clustering analysis, they have provided lower bounds for the critical density of a Poisson RGG in an $m$-dimensional Euclidean space.

However, to the best of our knowledge, neither the problem of setting the better recurrence threshold for the analysis of time series with non-uniform probability distribution, nor the problem of identifying metastability in recurrence networks, have been fully addressed.

When analyzing dynamical systems showing metastability, a different selection of recurrence threshold could reveal different structures associated to different time scales in the dynamics. A method for selecting an adequate recurrence threshold, that provides good estimations of the network's properties and a better understanding of the dynamical system, was still needed.


\section{\label{sec:Method}Method}

Our method for identification of metastable states in real-wold time series is divided into three parts. The first consists on the construction of the state space containing all the dynamical information of the dynamical system underlying a given time series. For this task, as mentioned in Section \ref{sec:Background}, there are multiple approaches. However, we use a methodology which procures the stability of the recurrences structure in state space.

Once the state space has been constructed, in order to perform the recurrence analysis of the time series, we set an appropriate recurrence threshold. The recurrence network computed with this recurrence threshold should capture all the dynamical information of the system. This means that every module on it must correspond to a metastable state in the time series.

Finally, we identify metastable states and transition region in the time series. For this, we use the method of Sarich et al.~\cite{Sari12, Sari12b, Bruc13, Djur12} to identify the modular structure in the recurrence network.


\subsection{\label{sec:StateSpace}Part I: State space construction}

To reconstruct the state space using the time delay embedding method, we need to set the embedding delay and the embedding dimension. The methodology by which we set these parameters is summarized in \ref{app:Algorithms}, Algorithm \ref{alg:alg1}. It is based in the analysis of two RQA measurements: entropy and recurrence rate.

The recurrence rate, $RR(\varepsilon)$, is a RQA measurement that indicates the density of recurrence points in a recurrence plot $R_{ij}(\varepsilon)$~\cite{Zbil92, Webb95}. When a time series has time points $N \to \infty$, it is the probability of recurrence of a state to its $\varepsilon$-neigborhood. In a recurrence network, it indicates a node's contribution to the relative frequency of edges~\cite{Donn11}. The recurrence rate is given by
\begin{equation}
RR(\varepsilon) = \frac{1}{N^2} \sum_{i,j=1}^N R_{ij}(\varepsilon)
\label{eq:recRate}
\end{equation}
Higher recurrence rate values indicate that the nodes in the recurrence network are more connected, or that a larger number of state space vectors fall inside a same state space neighborhood.

The Shannon entropy, $S(\varepsilon)$, indicates the complexity of the deterministic structure of a system and is expected to increase in presence of chaotic behavior. Lower entropy values indicate less time intervals with similar evolution in a time series. For a recurrence plot, it indicates the probability to find a diagonal line of length $l$ in $R_{ij}(\varepsilon)$~\cite{Raba07, Mar07}. Thus, it measures the complexity of a recurrence plot with respect to its diagonal lines. In terms of the diagonal lines of a recurrence plot, it is given by
\begin{equation}
S(\varepsilon) = -\sum_{l_{min}}^{N^*} p(l) \log p(l)
\label{eq:shannonEntropy}
\end{equation}
In this expression, $p(l)$ is the probability of finding a diagonal line of length $l$ in $R_{ij}(\varepsilon)$ and is given by $p(l)= P(l)/N_l$, $N_l= \sum_{l\geq l_{min}}P(l)$ and $P(l)=P(\varepsilon, l)$ is the histogram of diagonal lines. The length $l_{min}$ is a lower boundary for the diagonal lines that guarantees that all diagonal lines formed by the tangential motion of the state space trajectory will be excluded and also helps removing the effects of noise~\cite{Mar07}. We set this boundary as $l_{min}^* = \left[ \sum_{l=0}^{N^*} l P(l) \right]/\left[ \sum_{l=0} P(l) \right]$.

We select the embedding parameters as those providing simultaneous local minima in Shannon entropy and in recurrence rate, and having the largest Shannon entropy. We suggest that these restrictions construct a state space in which few nodes are neighbors but provide the maximum structure in the associated recurrence network.


\subsection{\label{sec:recurrenceAnalysisPart}Part II: Setting an appropriate recurrence threshold}

Small variations in the recurrence threshold can lead to very different modular structure in its associated recurrence network. This way, selecting an inadequate recurrence threshold can hide important dynamical structure in our data. We suggest that the recurrence threshold that better describes the dynamics of our data, lies in a region of values producing recurrence networks with similar modular structures. Since this region of values depends on the distribution of our time series data, which might not be uniform, we follow a self-adaptive methodology to set the recurrence threshold.

To capture the structure of the state space, we analyze the modular structure of recurrence networks in a filtration defined by the recurrence threshold. We suggest that an adequate recurrence threshold belongs to the subset of values in the filtration for which the modular structures of their associated recurrence networks are the most similar. The similarity in modular structure depends on the number and size of the modules identified in every recurrence network.


\subsubsection{\label{sec:SetRT}Constructing a set of recurrence networks}

The initial step for setting an adequate recurrence threshold consists on constructing a filtration defined by a set of recurrence thresholds, $\{ \varepsilon_{\nu} \}$. We construct the recurrence networks associated to the values in the filtration with the state space vectors reconstructed from the time series as described in Section~\ref{sec:StateSpace}. The similarity in modular structure of these networks will later help us compute a final recurrence threshold that captures the important dynamical information of the system underlying the time series data.

We want the filtration to span a wide range of values so that we can see different structures of the reconstructed state space in the associated recurrence networks. This way, we compute the initial recurrence threshold, $\varepsilon_0$, as the 95th percentile statistic of the distances between state space vectors. The smaller scale information we want to analyze is the one visible when the recurrence threshold is set as the 50th percentile statistic of the distances between state space vectors, denoted by $\varepsilon_f$. To estimate these distance values, we take 100 samples of the state space vectors, with length $n^*$ equal to $80\%N^*$.

Taking $\varepsilon_0$ and $\varepsilon_f$ as reference, we obtain the value of the recurrence thresholds in the filtration $\{ \varepsilon_{\nu} \}$, where $\nu = [0,\nu_f]$ and $\nu_f =14$. The $\nu$-th element of $\{ \varepsilon_{\nu} \}$ is thus given by
\begin{equation}
\varepsilon_{\nu} = \varepsilon_0+\nu \left( \frac{\varepsilon_f - \varepsilon_0}{\nu_f} \right) \label{eq:sequenceRecurrenceThreshold}
\end{equation}

The range and number of thresholds in the set could be modified according to information of the distribution of the state space vectors. In particular, if the data is uniformly distributed, the initial threshold could be given in terms of the standard deviation, as proposed by Marwan et al.~\cite{Mar07}. For multidimensional time series, we propose to use the largest standard deviation. In that case, we suggest the filtration to be given by $\varepsilon_{\nu} = \left(1.5-0.1 \nu \right) \varepsilon_0$.

Finally, we construct the associated recurrence network, $G_{\nu} = G(\varepsilon_{\nu})$, for every recurrence threshold in the filtration. This set of recurrence networks is denoted by $\{ G_{\nu} \}$.


\subsubsection{\label{sec:clusteringAnalysisPart}Finding a subset of recurrence networks with similar modular structure and setting a final recurrence threshold}

Every recurrence network in $\{ G_{\nu} \}$ may have a different modular structure. We analyze the number and size of the modules in every network with the aim of finding a similar subset.

The problem of finding modules, or clusters, in complex networks has been approached in several ways and many clustering algorithms exist for this purpose~\cite{Bruc13}. However, we use the algorithm of Sarich et al.~\cite{Sari12} because it is specifically developed for the case in which a system shows metastability. This algorithm is based on the spectral analysis of random walks on modular networks. It identifies modules as the metastable states in the random walker and transition regions composed by the nodes that do not belong to any metastable state. In computational terms, this algorithm scales linearly with the size of the network, making it also useful for analyzing large networks. In order to keep the sum of nodes assigned to all modules equal to $N^*$, we assign all nodes identified as part of the transition region to an additional module. This method does not work for disconnected or fully connected networks. This has been taken into account for defining the recurrence threshold values in the filtration.

The different modular structure in every recurrence network in $\{ G_{\nu} \}$ can be represented with a flow diagram called the Sankey diagram. In a Sankey diagram, every network is represented as a column and every column is divided into blocks. The number of blocks in a column represents the number of modules identified in a network. The size of every block in a column is determined by the number of nodes every module contains. Let $G_{\nu}$ and $G_{\nu+1}$ be two consecutive (recurrence) networks. Then, if a group of nodes initially assigned to module $A$ in $G_{\nu}$ is assigned to module $B$ in $G_{\nu+1}$, this \textit{flow} will be represented as an arrow in the Sankey diagram, with a thickness determined by the number of nodes \textit{flowing}. For more details, see \ref{app:sankey}.

\textbf{Similarity in number of modules.} The first similarity requirement on $\{ G_{\nu} \}$ is to have the same number of modules. The subset of recurrence networks satisfying this restriction is denoted by $\{ G_{\nu} \}^-$ and $\{ \varepsilon_{\nu} \}^-$ is the subset of recurrence thresholds in the filtration generating these networks.

Let $G_{\mu} \in \{ G_{\nu} \}$ have $C(G_{\mu})$ modules. Then, this network will satisfy the restriction on similarity in number of modules if, given three consecutive networks $G_{\mu-1},\,G_{\mu},\,G_{\mu+1} \in \{ G_{\nu} \}$, the following holds
\begin{eqnarray}
C(G_{\mu+1}) - C(G_{\mu}) &= 0 \nonumber \\
C(G_{\mu}) - C(G_{\mu-1}) &= 0 \nonumber \\
C(G_{\mu}) &> 1
\label{eq:numberOfClustersTolerance}
\end{eqnarray}

\textbf{Similarity in size of modules.} The next similarity requirement is applied to $\{ G_{\nu} \}^-$. It consists on asking these recurrence networks to have modules of similar size. The degree of similarity is expressed by a tolerance value, $\chi_j \in [ \chi_0, \chi^* ]$. The subset of networks satisfying this restriction with degree of similarity $\chi_j$ is denoted by $\{ G_{\nu} \}^{\chi_j}$. The subset of recurrence thresholds producing $\{ G_{\nu} \}^{\chi_j}$ is denoted by $\{ \varepsilon_{\nu} \}^{\chi_j}$.

Let $C_k(G_{\lambda})$ be the $k$-th module of $G_{\lambda} \in \{ G_{\nu} \}^-$, and $\vert C_k(G_{\lambda}) \vert$ the number of nodes in such module. Then, the size of the $k$-th module in a pair of consecutive recurrence networks $G_{\lambda}, G_{\lambda+1} \in \{ G_{\nu} \}^-$ varies less than $\chi_j$ if
\begin{equation}
\vert C_k(G_{\lambda+1}) \vert -\vert C_k(G_{\lambda}) \vert < \chi_j
\label{eq:sizeOfClustersTolerance}
\end{equation}

We define the tolerance value in terms of the number of nodes in the recurrence networks, $N^*$. Initially, we say that two modules have similar size if the number of nodes they contain is different in no more than ten percent of $N^*$. This means that $\chi_0=0.1N^*$. By decreasing the tolerance value, we strengthen the condition of similarity between modules. We say that the maximum similarity is reached when the number of nodes in two modules is different in no more than one percent of $N^*$, which means that $\chi^*= 0.01N^*$. We define a ten steps procedure, where the tolerance value for each step is given by
\begin{equation}
\chi_j = \chi_0(1- j/10) \mbox{, for $j=[0,9]$}  \label{eq:sizeOfClustersToleranceChi}
\end{equation}

If the subset of recurrence networks satisfying the maximum decrease of tolerance is not empty, it is denoted by $\{ G_{\nu} \}^* = \{ G_{\nu} \}^{\chi^*}$. Then, the subset of recurrence thresholds producing these networks is denoted by $\{ \varepsilon_{\nu} \}^* = \{ \varepsilon_{\nu} \}^{\chi^*}$. However, it is possible that no subset of $\{ G_{\nu} \}^-$ satisfies the maximum tolerance decrease and that $\{ G_{\nu} \}^{\chi_{j}}$ is empty for a certain tolerance $\chi_0 \geq \chi_! > \chi^*$. In this case, we define $\{ G_{\nu} \}^* = \{ G_{\nu} \}^{\chi_!}$ and $\{ \varepsilon_{\nu} \}^* = \{ \varepsilon_{\nu} \}^{\chi_!}$.

Finally, we set the \textit{final recurrence threshold}, $\varepsilon^*$, as the smallest recurrence threshold in $\{ \varepsilon_{\nu} \}^*$.


\subsection{\label{sec:Classification}Part III: Identifying metastable states in the time series}

Once that the final recurrence threshold $\varepsilon^*$ has been set, we generate the recurrence network associated to it, $G_*=G(\varepsilon^*)$. The analysis of the modular structure of this network will lead to the identification of metastable states (and transition region) in the time series.

In this paper, for simplicity, if node $i$ of $G_*$ has been assigned to a specific module $C_k$, the data point $u_i$ in the first component of state space vector $\overrightarrow{x}_i$ is assigned to the $k$-th metastable state.

This metastable state assignment approach is na\"ive because, when using the time delay embedding method to construct the state space, every data point appears in a different number of state space vectors. Let $M(u_i)$ denote the number of state space vectors in which data point $u_i$ appears, $\tau$ and $m$ be the embedding parameters, and $\alpha$ be an integer such that $0 \leq \alpha < m-1$. Then, $M(u_i) = \alpha + 1$ if $\alpha \tau \leq i \leq (\alpha + 1)\tau$ or if $N-(\alpha+1)\tau < i \leq N-\alpha\tau$, and $M(u_i) = m$ for any other data point.

Alternatively, the metastable state a data point $u_i$ is assigned to, could be determined by its \textit{dominant module}. This means, the module to which most of the state space vectors in which $u_i$ is have been assigned to. 
Let $\mathbf{x}_i$ be a state space vector of which data point $u_i$ is a component. It $\mathbf{x}_i$ is assigned to more than one module the same number of times, then there is no dominant module for $u_i$ and one could consider it part of the transition region.


\section{\label{sec:Results}Examples}

To illustrate the ability of our methodology to identify metastable states in complex time series, we present and analyze two cases. As the results of these analysis suggest, our methodology is able to identify metastable states even in complex time series with noise and missing data.

First, we analyze the one-dimensional time series describing the motion of a particle under the gradient of a double well potential and a random force. This is one of the simplest systems showing metastability.

Then we analyze a one-dimensional real-world time series containing the average daily temperatures in Berlin-Tempelhof from June 12th, 1936 to January 9th, 2008. This time series is likely to have trends and has several missing measurement points during some periods of time.

\subsection{\label{sec:Twp}Double well potential}

\begin{figure}[h!]
\centering 
  \subfigure[][ Original time series.]{\includegraphics[height = 0.28\paperwidth]{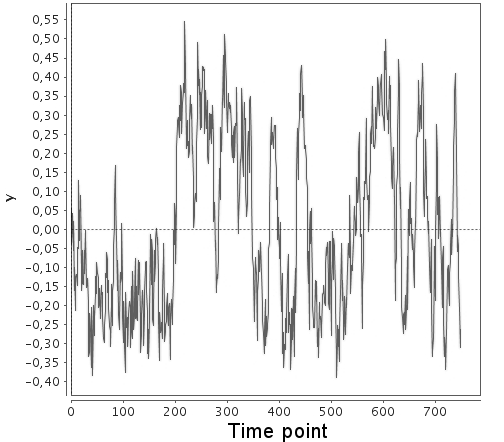} \label{fig:doubleWellTrajectory}}\hspace{0.1cm}
  \subfigure[][ Metastable states identified in the time series.]{\includegraphics[height = 0.28\paperwidth]{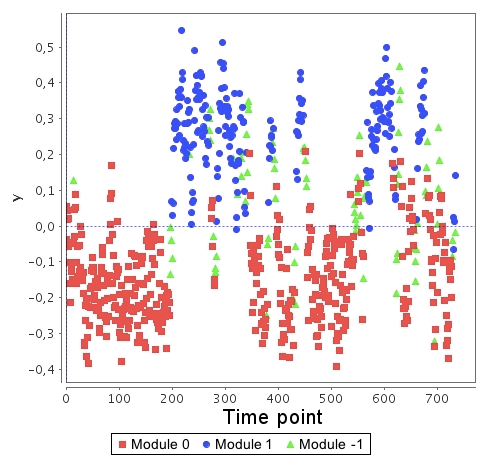} \label{fig:doubleWellClassified}}
\caption{\label{fig:doubleWellPotential}Time series for a particle in a double well potential. Fig.~\ref{fig:doubleWellTrajectory} shows the original time series, resulting from normalizing the time series computed by integration of the double well potential Langevin equations and sampling every 10 data points. For the integration we use an Euler Maruyama integrator with lag time $\lambda = 0.001$, 7500 iterations, initial positions $ q_{init} = (0,1)$ and temperature $T=100$. The color code in Fig.~\ref{fig:doubleWellClassified} shows the different metastable states identified on the original time series. Modules 0 and 1 indicate the two metastable states. The data points assigned corresponding to the transition region are represented by module -1. The embedding parameters are set to $\tau = 7$ and $m = 2$. The recurrence threshold is set to $\varepsilon^* \approx 0.29$.}
\end{figure}

Our first example is the time series describing the motion of a particle in a heat bath with temperature $T$, under the gradient of a double well potential and a random force (see Fig.~\ref{fig:doubleWellTrajectory})

This model, proposed by Kramer in 1949~\cite{Kram49} during his studies on chemical reactions, is one of the first models for metastability. It can be described by
\begin{equation}
dX_t = -\nabla V(x) dt + \sqrt{2\epsilon} dB_t \label{eq:twpEquation}
\end{equation}
Where $B_t$ is a Brownian motion, $\nu>0$ is a friction parameter and $\epsilon = \nu T$.
The potential is given by $V(x) = (x^2-a^2)^2$. It has two local minima at $x_1=a$ and $x_2=-a$. For this example we set $a=1$. The trap depth difference between the potential wells,  $\Delta V$, controls how metastable the system is.

Our one-dimensional time series, shown in Fig.~\ref{fig:doubleWellTrajectory}, results from integrating the double well potential's Langevin dynamical equations. For this, we use the Euler Maruyama integrator with lag time $\lambda = 0.001$, 7500 iterations, initial positions $q_{init} = \left(0,1\right)$ and temperature $T = 100K$. Additionally, we sample this time series every 10 time points. Therefore, the length of this time series is $N=750$ data points.

In this time series, we expect to find two metastable states, corresponding to each of the potential wells, and a transition region, indicating the moments of transition between potential wells. 

For the analysis of this time series we set $\tau=7$ and $m=2$, so we reconstruct $N^*=736$ state space vectors. These parameters were determined as explained in \ref{app:Algorithms}, algorithm \ref{alg:alg1}, taking $\tau_0=1, \tau_F = 10, m_0=2$ and $m_F=8$.

The results of our analysis are shown in the Sankey diagram of Fig.~\ref{fig:dwpSankey}. As this figure shows, the size of the metastable modules (and transition region) satisfies the maximum similarity restrictions (Eq.~\ref{eq:sizeOfClustersTolerance} with $\chi^*=0.01N^* \approx 7$) for recurrence thresholds $\varepsilon \in [\varepsilon_{4},\varepsilon_{6}]$. The modular structure analysis of every recurrence network are used to set the final recurrence threshold, $\varepsilon^* \approx 0.29$.

Finally, the modules in the final recurrence network, computed with $\varepsilon^*$, indicate the metastable states in the time series. These are shown in Fig.~\ref{fig:doubleWellClassified}, where modules 0 and 1 can be associated to the two expected metastable states, one for every potential well. The nodes assigned to the transition region are represented in this figure as module -1.


\subsection{\label{sec:Weather}Weather data}

\begin{figure}[h!]
\centering 
  \subfigure[][ Original time series.]{\includegraphics[width = 0.45\textwidth]{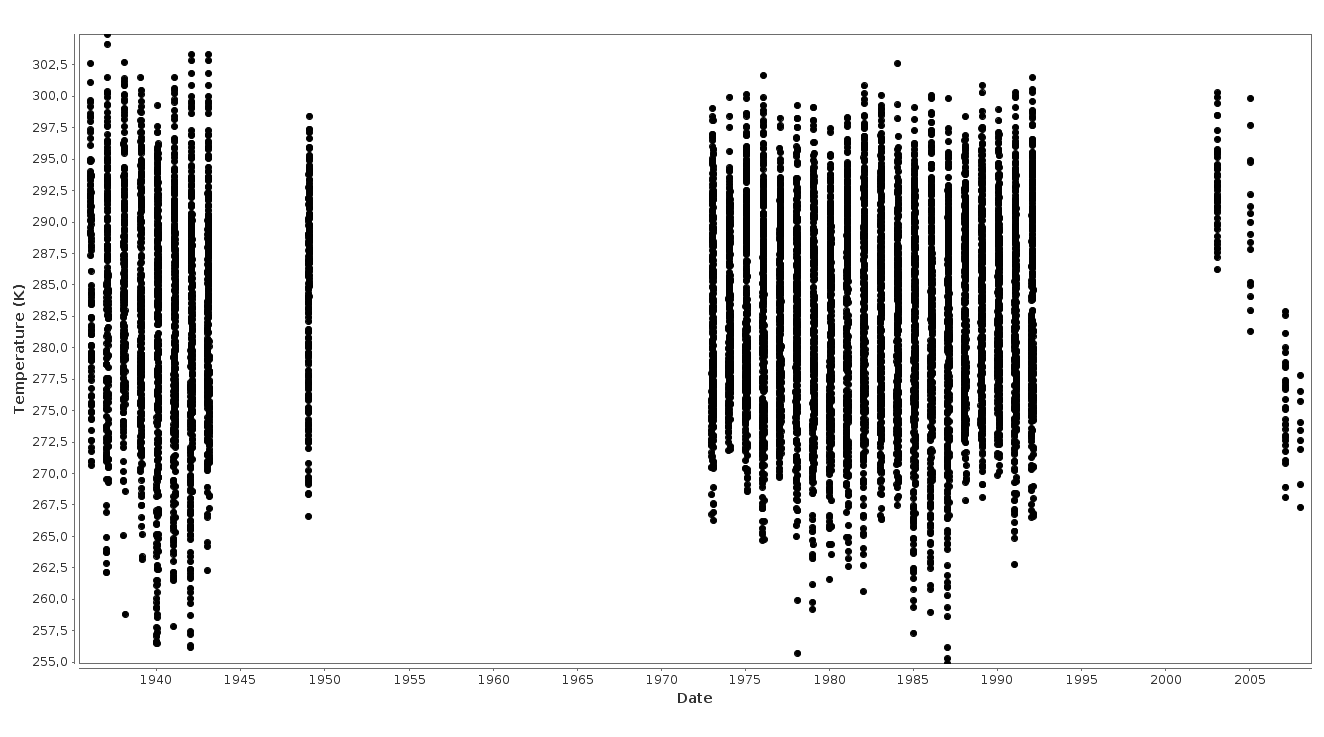}\label{fig:berlinWeatherFull}}\hspace{0.1cm}
  \subfigure[][ \textit{Merged} time series.]{\includegraphics[width = 0.45\textwidth]{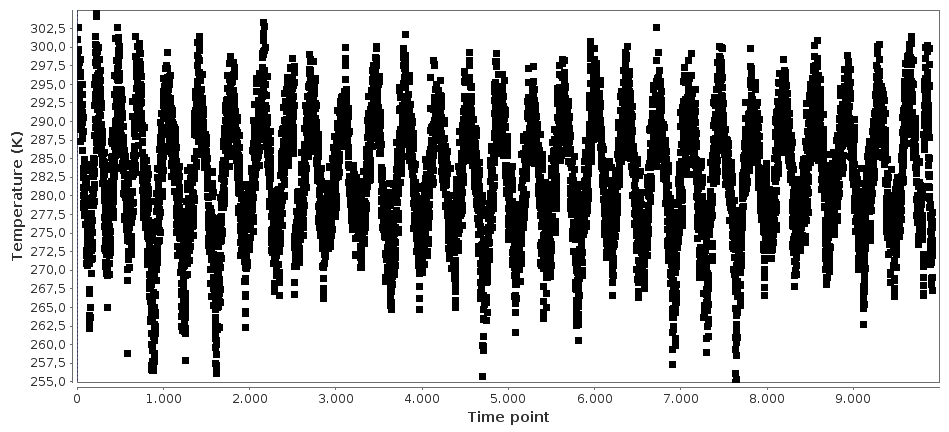}\label{fig:berlinWeatherCoarse}}
\caption{\label{fig:berlinWeather}Daily average temperatures, in $K$, in Berlin-Tempelhof weather station from June 12, 1936 to January 9, 2008. Fig. \ref{fig:berlinWeatherFull} shows the original time series. Measurements are irregularly taken before August 31, 1939 and measuring techniques previous to 1943 are not provided. Periods without measurements are due to historical or technical reasons. Fig. \ref{fig:berlinWeatherCoarse} shows the \textit{merged} time series, obtained from the previous time series by ignoring the time points in which no measurements were taken.}
\end{figure}

Our second example corresponds to the observations of the average daily temperatures in Berlin-Tempelhof weather station, from June 12, 1936 to January 9, 2008 (see Fig.~\ref{fig:berlinWeatherFull}). This time series is taken from the Rimfrost database~\cite{rimfrost}, which collects information from the German Weather Service~\cite{dwd} (\textit{Deutscher Wetterdienst}) and the NASA Goddard Institute for Space Studies~\cite{nasagiss} (NASA-GISS). It includes several periods without measurements. We will refer to this, as the \textit{complete} time series.

Our analysis is performed in the time series shown in Fig.~\ref{fig:berlinWeatherCoarse} and so called \textit{merged} time series. This is obtained by ignoring the periods of time without measurements in the original time series. To the best of our knowledge, the time series data analyzed in this section has not been previously analyzed in any similar fashion. However statistical analysis and interpretations of such analysis have been performed~\cite{stadtEntw}.

\begin{figure*}
\centering 
  \subfigure[][ Data points from January 1, 1937 to December 31, 1938.]{\includegraphics[height = 0.18\paperheight]{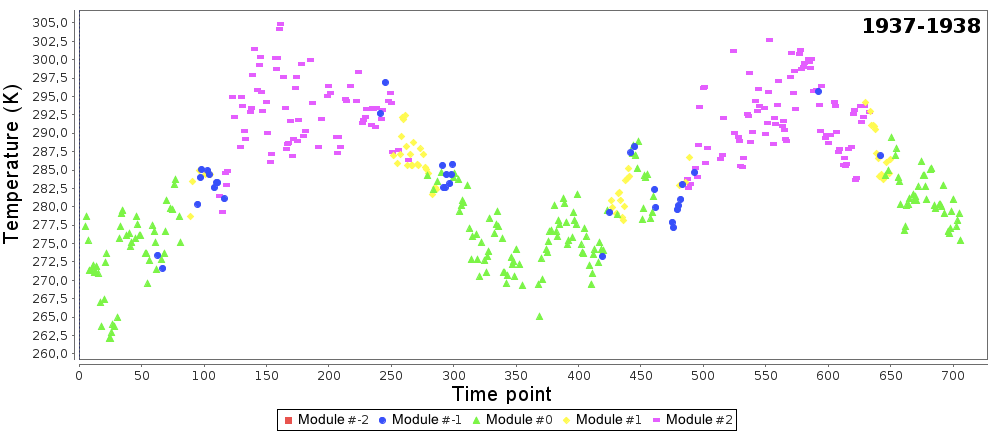}\label{fig:weather3738}}\\
  \subfigure[][ Data points from January 1, 1942 to December 31, 1943.]{\includegraphics[height = 0.18\paperheight]{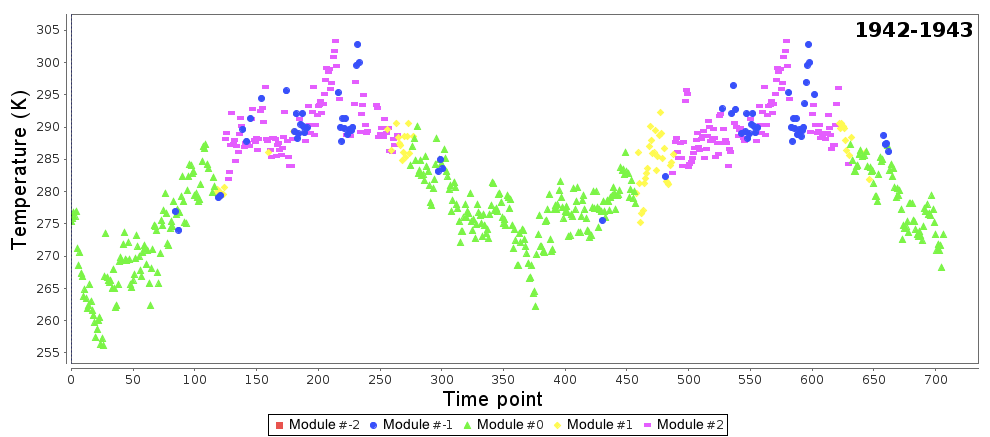}\label{fig:weather4243}}\\
  \subfigure[][ Data points from January 1, 1991 to December 31, 1992.]{\includegraphics[height = 0.18\paperheight]{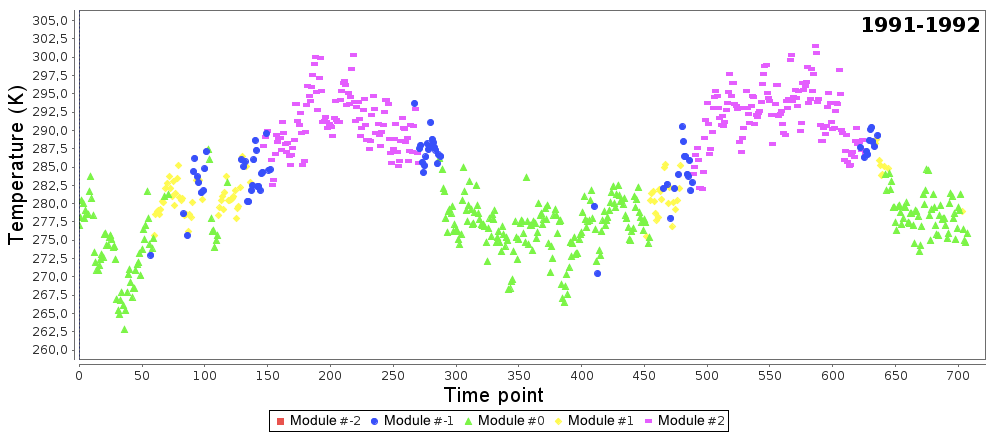}\label{fig:weather9192}}\\
  \caption{The color code in these figures represents the different metastable states (modules 0, 1 and 2) and transition region (module -1) identified (using $\varepsilon^* \approx 0.52$, $\tau = 3$ and $m = 7$) in different periods of the time series containing the daily average temperatures measured in Berlin-Tempelhof from June 12, 1936 to January 9, 2008. Time points with missing temperature measurements are assigned temperature $T=0^{\circ}K$ and included in module -2 (not shown in the plots). For convenience, this module is not shown in the plots. In Fig. \ref{fig:weather3738}, temperature measurements are irregularly taken and there is no information about the measuring technique. In Fig. \ref{fig:weather4243} and Fig. \ref{fig:weather9192}, the measuring technique is specified and there are few or none missing measurements.}
   \label{fig:segmentsWeather}
    \end{figure*}

To simplify the computations, we sample the merged time series every 14 time points to produce a \textit{coarse} time series. In the periods in which measurements are regularly taken, this sampling rate corresponds to taking the daily temperature every second week and therefore we suggest that season transitions could be sufficiently represented. Evidently, this is not the case in the periods in which measurements are irregular and we can not guarantee the appropriate representation of seasons. For this reason, we use the \textit{coarse} time series to compute a final recurrence threshold but later identify metastable states in different sections of the merged time series.

For our analysis we use the \textit{coarse} time series. By adding the missing time points to the analyzed merged time series, assigning them a temperature $T=0K$ and indicating them by module -2, we describe the behavior of the complete time series. We select embedding parameters $\tau = 3$ and $m = 7$, and final recurrence threshold, $\varepsilon^* \approx 0.52$.

We analyze three segments of the merged time series, corresponding to the following periods of time: (a) from January 1, 1937 to December 31, 1938, (b) from January 1, 1942 to December 31, 1943, and (c) from January 1, 1991 to December 31, 1992. In these, we expect to identify yearly seasons and the transit between them.

The results of analyzing the first period of time are shown in Fig.~\ref{fig:weather3738}. This period has several missing measurements$\,$--$\,$around 30$\%$ of the time points$\,$--$\,$and there is no information about the way in which measurements were taken. The results of analyzing the second period of time are shown in Fig.~\ref{fig:weather4243}. This period does not have many missing measurements: less than 1$\%$ of the time points. Temperature measurements in this period are less disperse than in the previous period, which suggests a more systematic measuring technique. Results for the third period are shown in Fig.~\ref{fig:weather9192}. This period does not have missing measurements. Temperature measurements during this period were obtained more systematically.

In all time series we identify one metastable state corresponding to a colder season, module 0, which lasts around six months. Another metastable state, module 2, can be associated to the warmer season. Module 1 can be associated to the mild seasons between the colder and warmer. The time points assigned to the transition region in the recurrence network are indicated by module -1. This module does not seem to correspond to a particular season in Fig.~\ref{fig:weather4243}, but in Fig.~\ref{fig:weather3738} and Fig.~\ref{fig:weather9192}, seems to also correspond to the period of time between warmer and colder seasons. This location of Module -1 might result from the large amount of missing measurement points in the time series shown in Fig.~\ref{fig:weather3738} and Fig.~\ref{fig:weather4243}. A suggestion to improve the identification of metastable states in this time series is to analyze it with different recurrence threshold and embedding parameters, specific for these data and not for the \textit{coarse} time series. 


\section{\label{sec:Robustness}Robustness}

In this section we measure how robust is our method for identifying metastable states in two scenarios: when a percentage of noise is added and when a percentage of time points is removed from a time series. We define robustness as the similarity between the metastable states identified in the \textit{original} time series and in the time series \textit{modified} by the noise or data points removal.

For these analysis we take as example a double well potential time series. The parameters we use to analyze all double well potential noisy time series are $\varepsilon^* \approx 0.41$, $\tau = 1$ and $m = 5$.

The similarity between these two time series is measured with the Adjusted Rand Index~\cite{ARI} (ARI), developed by Hubert and Arabie in 1985. This index measures the agreement between two partitions. It ranges from 0, when the partitions are not similar at all, to 1, when the partitions are equivalent. It can be computed even if the number of modules in the two partitions compared is different. In our case, a partition is given by the assignment of every state space vector to a module. For the expression of this index, see \ref{app:ari}.

Since we analyze the modular structure of recurrence networks with the algorithm developed by Sarich et al.~\cite{Sari12}, we use the adaption to this index proposed by Hueffner et al.~\cite{Bruc13}. This modification accounts for the division of the networks into modular and transition regions. It assigns every state space vector identified as part of the transition region to an independent module in order to create a full partition. The ARI can then be measured considering only the modules or the modules together with the transition region.


\subsection{\label{sec:noise}Noisy time series}

The robustness of our method is measured as the difference in results between a time series and a noisy time series computed by adding noise to the former one. Noise is defined as a percentage of the amplitude of the time series. Our results suggest that our method is robust to noise with up to 7$\%$ the amplitude of the original time series.

A noisy time series is created by adding Gaussian white noise (mean equal to zero and standard deviation equal to one) to the time series. The amplitude of the noise is equal to a multiple, $\mu$, of the amplitude of the original time series. We vary the amplitude of noise, $\mu$, from 0 to 20 in intervals $\Delta \mu = 1$.

In order to get rid of the bias produced by the selection of noise, for every increase in $\mu$ we analyze 50 different noisy time series. The final ARI is the average of the measurements for every time series with the same amplitude of noise and therefore we introduce error bars showing a confidence interval of 90$\%$.

\begin{figure}[h!]
\centering 
  \subfigure[][ ARI measured only in the modules.]{\includegraphics[height = 0.28\paperwidth]{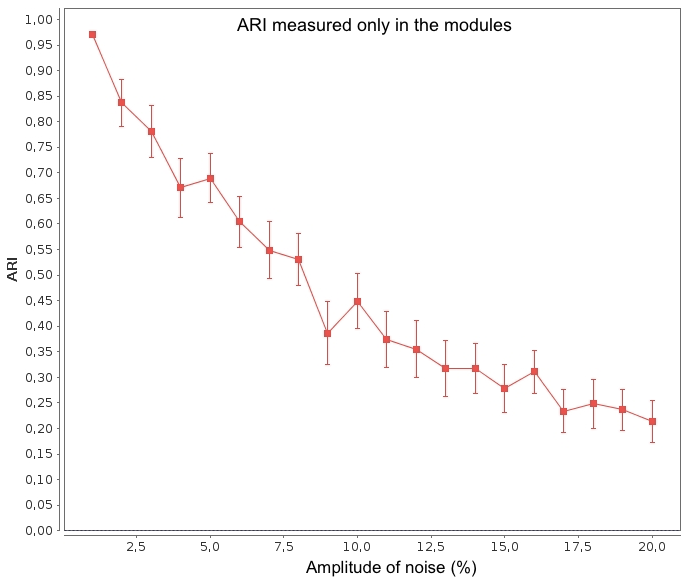}\label{fig:13a}}
  \subfigure[][ ARI measured in the modules and transition region.]{\includegraphics[height = 0.28\paperwidth]{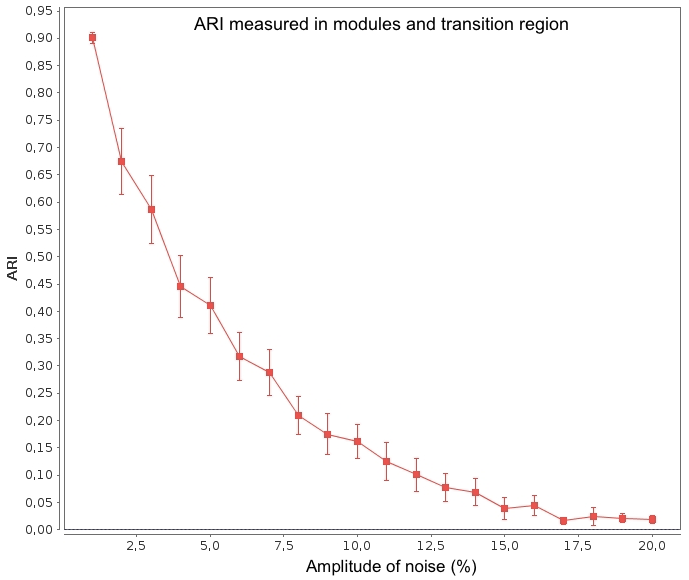}\label{fig:13b}}
\caption{\label{fig:testNoisePercentage}Robustness to \textbf{noise}. Similarity between the metastable states identified in a time series and in its modified version, where noise has been added. The noisy time series is created by adding white Gaussian noise with amplitude equal to a percentage, $\mu$, of the amplitude of the original time series. Similarity is measured with the Adjusted Rand Index (ARI) in two ways: considering only the modules [Fig.~\ref{fig:13a}] and considering modules and transition region [Fig.~\ref{fig:13b}]. Error bars show confidence interval of 90$\%$.  The state space for every time series is reconstructed using $\tau = 1$ and $m = 5$. The recurrence threshold used to build the recurrence network is set to $\varepsilon^* \approx 0.41$.}
\end{figure}

Figure~\ref{fig:testNoisePercentage} shows the results of our analysis. These indicate that the ARI is around 0.6 for $\mu \lesssim 7$, when measured only in the modules. When measured considering modules and transition region, the amplitudes of noise for which ARI is around 0.6, fall down to $\mu \lesssim 3$.

As mentioned by Zbilut in 1992~\cite{Zbil92}, having noise in a time series inflates the embedding dimension needed to reconstruct the state space. Therefore, if every noisy time series were analyzed with different recurrence threshold and embedding parameters, we would expect that the similarity between original and noisy time series would hold for noise with larger amplitude.


\subsection{\label{sec:missing}Removing data points}

Another typical feature of real-world time series is having measurements irregularly taken. We understand these irregularities as removing a percentage of measurement points, randomly distributed, form a time series containing measurements regularly taken. 

Therefore, we produce a time series with regularly spaced measurements. Then, we modify this time series by assigning a ``null" value to a percentage of randomly distributed data points. We do not ignore time points but rather assign them a new value, in order to keep equal the length of the original and the modified time series. We vary the percentage of time points being removed from 0$\%$ to 19$\%$, in intervals of 1$\%$. Again, in order to get rid of the bias produced by the selection of data points to remove, we analyze 50 different time series with a same percentage of data points removed.

\begin{figure}[h!]
\centering 
  \subfigure[][ ARI measured only in the modules.]{\includegraphics[height = 0.28\paperwidth]{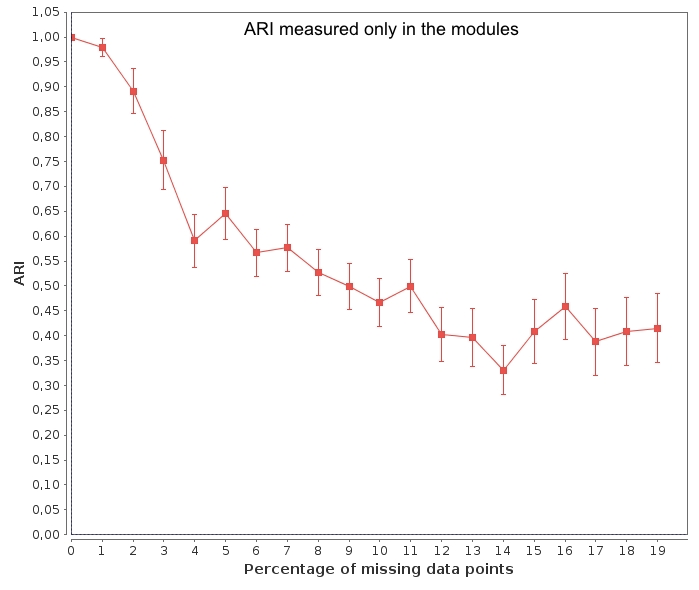}\label{fig:14a}}
  \subfigure[][ ARI measured in the modules and transition region.]{\includegraphics[height = 0.28\paperwidth]{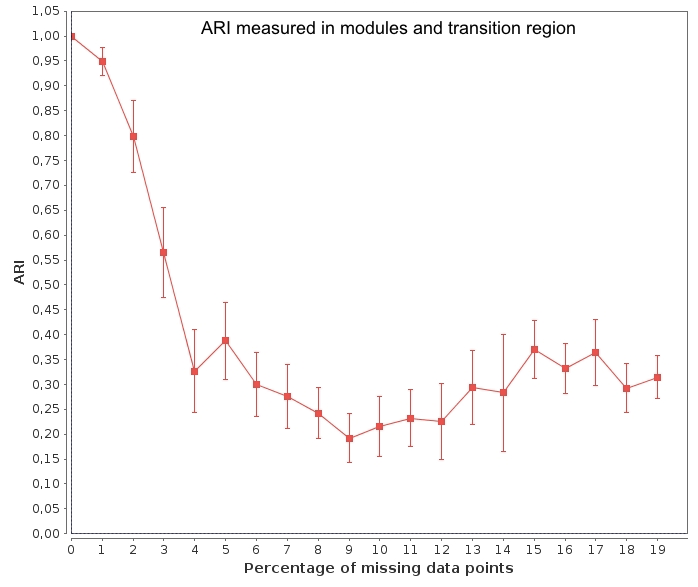}\label{fig:14b}}
\caption{\label{fig:missingTest1}Robustness to \textbf{missing data points}. Similarity between the metastable states identified in a time series and in its modified version, where a percentage of randomly distributed data points has been removed. Similarity is measured with the Adjusted Rand Index (ARI) in two ways: considering only the modules [Fig.~\ref{fig:14a}] and considering modules and transition region [Fig.~\ref{fig:14b}]. Error bars show confidence interval of 90$\%$. The state space for all time series is reconstructed using $\tau = 1$ and $m = 5$. The recurrence threshold used to build the recurrence network is set to $\varepsilon^* \approx 0.41$.}
\end{figure}

Similar to the results obtained when analyzing noise, Fig.~\ref{fig:missingTest1} shows that the ARI is around 0.6 for $\mu \lesssim 7$, when measured only in the modules. When measured considering modules and transition region, the amplitudes of noise for which ARI is around 0.6, fall down to $\mu \lesssim 3$. Therefore, we could interpret the missing data points as another case of noise, causing the inflation of the embedding dimension.

These results indicate that our method is robust even for time series with up to 19$\%$ of randomly distributed missing points.


\section{\label{sec:End}Conclusions}

In this paper, we present a self-adaptive method for the identification of metastable states in real-world time series based on recurrence networks analysis.

Our method uses particular statistical information of a given time series in order to produce a filtration defined by the recurrence threshold. The analysis of the modular structure of the recurrence networks associated to this filtration results in the classification of almost all data points in the time series into different metastable states or transition region.

For the reconstruction of the state space from a time series, necessary for any recurrence analysis, we use the delay mapping. The appropriate embedding parameters (delay and dimension) for this reconstruction, depend on the properties of the time series. Therefore, we also propose a methodology to set these parameters, which depends on the first simultaneous minima of entropy and recurrence rate in the recurrence plots associated to the filtration. The selection of these parameters, prior to the recurrence analysis, is still an open problem that we aim to approach in future work.

Additionally, we analyze the robustness of our method. This is done by measuring the similarity between the metastable states identified in a time series and to a modified version of the same: obtained by adding noise or removing data points. The similarity is measured in the partitions (into metastable states and transition region) of the associated recurrence networks with a modified Adjusted Rand Index~\cite{ARI} (ARI) that accounts for the existence of metastable states and transition region in the time series. Measuring the similarity between partitions in the metastable states only (excluding the transition region), we obtain that $ARI > 0.6$ even when up to 7$\%$ of data points is removed. For time series with noise, where the amplitude of noise is expressed as a percentage of the amplitude of the time series, $ARI > 0.6$ even for $\approx 7\%$ noise amplitude. 

The results from these analysis suggest that our method is an adequate tool for the identification of metastable states in complex time series, even in the presence of low percentages of noise and missing data points.


\section{Acknowledgements}

I. Vega is part of the International Max Planck Research School for Computational Biology and Scientific Computing and the Freie Universit\"at Berlin. The funding institutions had no involvement in the study design; collection, analysis and interpretation of the data; nor in the writing of the report or in the decision to submit the article for publication.


\section*{References}
\bibliographystyle{elsarticle-num} 
\bibliography{Vega_SAIMER_Arxiv}
\newpage
\appendix

\section{\label{app:ari}The adjusted rand index (ARI)}
Let us imagine $S=\{ O_1,...,O_N \} $, a set of objects. The number of combinations of pairs that are possible to make from set $S$ is $\binom{N}{2}$. Set $P=\{ p_1, p_2, ..., p_A\}$ and $Q=\{ q_1, q_2, ...,  q_B\}$ two partitions (or collections of subsets) of $S$ such that $\cup_{a=1}^A p_a = \cup_{b=1}^B q_b = S$, $p_a \cap p_{a'} = \emptyset$ for any $a \neq a'$, and $q_b \cap q_{b'} = \emptyset$ for any $b \neq b'$. If $t_{ab}$ represents the number of objects in $S$ that were classified in the $a$-th subset of $P$ and in the $b$-th subset of $Q$, then the ARI, as defined by Santos~\cite{ARI2}, can be expressed as the quotient $F_1/F_2$, where:
\begin{equation*}
F_1 = \binom{n}{2} \sum_{a=1}^{A} \sum_{b=1}^B \binom{t_{ab}}{2} - \sum_{a=1}^A \binom{t_{a \cdot}}{2} \sum_{b=1}^B \binom{t_{\cdot b}}{2}
\end{equation*}

\begin{equation*}
F_2 = \frac{1}{2} \binom{n}{2} \left[ \sum_{a=1}^A \binom{t_{a \cdot}}{2} + \sum_{b=1}^B \binom{t_{\cdot b}}{2} \right] - \sum_{a=1}^A \binom{t_{a \cdot}}{2} \sum_{b=1}^B \binom{t_{\cdot b}}{2}
\end{equation*}


\section{\label{app:Algorithms}Algorithms}


\begin{algorithm}[H]
\caption{\label{alg:alg1} \textsc{. Construct the state space}}
\begin{algorithmic}
\small
\State $\bullet$ \; \textsc{Normalize time series}
\vspace{2pt}
\State Normalize to the maximum the given time series, containing $N$ data points, in an interval going from zero to one (min-max normalization).
\vspace{6pt}
\State $\bullet$ \; \textsc{Compute RQA measurements for a set of embedding parameters}
\State \quad \textbf{for} $\tau = \tau_0$ \textbf{to} $\tau = \tau_F$ \textbf{do}
	\begin{itemize}[itemsep=0pt, parsep=0pt, labelindent=0pt,labelwidth=\widthof{\ref{last-item}},itemindent=0em,leftmargin=!]
	\item[] \textbf{for} $m =m_0$ \textbf{to} $m = m_F$ \textbf{do}
		\begin{itemize}
		\item[$\triangleright$] For embedding parameters $\tau$ and $m$, construct $N^* = N - (m-1) \tau$ state space vectors using the time delay embedding method (see Eq.~\ref{eq:phaseSpaceTrajectory}) from the normalized time series.
		\item[$\triangleright$] Compute recurrence threshold, $\varepsilon$, as explained in Section \ref{sec:recurrenceAnalysisPart}.
		\item[$\triangleright$] Compute its associated recurrence plot $R_{ij}(\varepsilon)$.
		\item[$\triangleright$] Compute $RR(\varepsilon)$ and $S(\varepsilon)$ for the associated recurrence plot, as given in Eqs.  \ref{eq:recRate} and \ref{eq:shannonEntropy} respectively.
		\end{itemize}
	\item[] \textbf{end for}
	\end{itemize}
\State \quad \textbf{end for}
\vspace{6pt}
\State $\bullet$ \; \textsc{Select embedding parameters}
\vspace{2pt}
\State Select $m$ and $\tau$ that first provide a simultaneous local minima in entropy and in recurrence rate, and that also give the highest entropy.
\end{algorithmic}
\end{algorithm}


\begin{algorithm}[H]
\caption{\label{alg:alg2} \textsc{. Set appropriate recurrence threshold}}
\begin{algorithmic}
\small
\vspace{6pt}
\State $\bullet$ \; \textsc{Construct set of recurrence networks}
\begin{itemize}[itemsep=0pt, parsep=0pt, labelindent=0pt,labelwidth=\widthof{\ref{last-item}},itemindent=0em,leftmargin=!]
\item[] \textbf{for} $\nu=0$ \textbf{to} $\nu=\nu_{f}$ \textbf{do}
	\begin{itemize}[itemsep=0pt, parsep=0pt, labelindent=0pt,labelwidth=\widthof{\ref{last-item}},itemindent=0em,leftmargin=!]
	\item[$\triangleright$] Compute recurrence threshold $\varepsilon_{\nu}$ according to Eq. \ref{eq:sequenceRecurrenceThreshold}
	\item[$\triangleright$] Compute associated recurrence plot $R(\varepsilon_{\nu})$ and recurrence network $G_{\nu} = G(\varepsilon_{\nu})$.
	\end{itemize}
\item[] \textbf{end for}
\item[] \textbf{return} Set of recurrence networks $\{ G_{\nu} \}$.
\end{itemize}
\vspace{6pt}
\State $\bullet$ \; \textsc{ Find modules in a set of recurrence networks}
\begin{itemize}[itemsep=0pt, parsep=0pt, labelindent=0pt,labelwidth=\widthof{\ref{last-item}},itemindent=0em,leftmargin=!]
\item[] \textbf{for} $\nu=0$ \textbf{to} $\nu=\nu_{f}$ \textbf{do}
	\begin{itemize}[itemsep=0pt, parsep=0pt, labelindent=0pt,labelwidth=\widthof{\ref{last-item}},itemindent=0em,leftmargin=!]
	\item[$\triangleright$] Perform modular structure analysis of associated recurrence network $G_{\nu}=G(\varepsilon_{\nu})$.
	\item[$\triangleright$] Compute number of modules, $C(G_{\nu})$, and number of nodes in each module, $\vert  C_k(G_{\nu}) \vert$, on $G_{\nu}$.
	\end{itemize}
\item[] \textbf{end for}
\item[] \textbf{return} Modular structure information for all networks in $\{G_{\nu}\}$: $\{C(G_{\nu})\}$ and $\{|C_k(G_{\nu})|\}$.
\end{itemize}
\vspace{6pt}
\State $\bullet$ \; \textsc{ Select subset of similar networks}
\begin{itemize}[itemsep=0pt, parsep=0pt, labelindent=0pt,labelwidth=\widthof{\ref{last-item}},itemindent=0em,leftmargin=!]
\item[$\triangleright$] Select subset of networks with the same number of modules, $\{ G_{\nu} \}^-$, satisfying Eq. \ref{eq:numberOfClustersTolerance}.
	\item[] \textbf{for} $\chi_j=\chi_0$ \textbf{to} $\chi_j=\chi^*$ (defined in Eq. \ref{eq:sizeOfClustersToleranceChi}) \textbf{do}
		\begin{itemize}[itemsep=0pt, parsep=0pt, labelindent=0pt,labelwidth=\widthof{\ref{last-item}},itemindent=0em,leftmargin=!]
		\item[] \textbf{for all} $\varepsilon_{\lambda} \in \{ \varepsilon_{\nu} \}^-$ \textbf{do}
			\begin{itemize}[itemsep=0pt, parsep=0pt, labelindent=0pt,labelwidth=\widthof{\ref{last-item}},itemindent=0em,leftmargin=!]
			\item[$\triangleright$] Select subset of networks in $\{ G_{\nu} \}^-$ with similar size of modules, i.e. satisfying Eq. \ref{eq:sizeOfClustersTolerance}.
			\end{itemize}
		\item[] \textbf{end for}
		\item[] \textbf{if} $\{ \varepsilon_{\nu} \}^{\chi_j} = \emptyset$ \textbf{then}
			\begin{itemize}[itemsep=0pt, parsep=0pt, labelindent=0pt,labelwidth=\widthof{\ref{last-item}},itemindent=0em,leftmargin=!]
			\item[] $\chi_{!}=\chi_{(j-1)}$ and $\{ \varepsilon_{\nu} \}^{*} = \{ \varepsilon_{\nu} \}^{\chi_{!}}$
			\end{itemize}
		\item[] \textbf{else} $\{ \varepsilon_{\nu} \}^{*} = \{ \varepsilon_{\nu} \}^{\chi^*}$
		\item[] \textbf{end if}
		\end{itemize}
	\item[] \textbf{end for}
	\item[] \textbf{return} Subset  $\{ \varepsilon_{\nu} \}^{*}$.
\item[$\triangleright$] Set the final recurrence threshold, $\varepsilon^*$, as the minimum threshold in $\{ \varepsilon_{\nu} \}^*$.
\end{itemize}
\end{algorithmic}
\end{algorithm}


\begin{algorithm}[H]
\caption{\label{alg:alg3} \textsc{. Identify metastable states}}
\begin{algorithmic}
\small
\vspace{6pt}
\State $\bullet$ \; \textsc{Identify Metastable States}
	\begin{itemize}[itemsep=0pt, parsep=0pt, labelindent=0pt,labelwidth=\widthof{\ref{last-item}},itemindent=0em,leftmargin=!]
	\item[$\triangleright$] Perform modular structure analysis of recurrence network $G_*=G(\varepsilon^*)$.
	\item[$\triangleright$] Classify time points into different metastable states, according to any of the two methods proposed in Section \ref{sec:Classification}: depending on the first state vector component or on the \textit{dominant module}.
	\end{itemize}
\end{algorithmic}
\end{algorithm}


\section{\label{app:sankey}Sankey diagram for two well potential time series analysis.}

A Sankey diagram is a visual tool that shows the number and size of modules in a network, as well as the changes in these modules when a parameter is modified.

In these diagrams, each network is represented as a column, the number of modules in a network is represented by the number of sections in a column and the size of each section corresponds to the number of nodes in each module in the corresponding network. The amount of nodes whose correspondence to a module varies from one network to another, is represented as a flux between columns, and the width of such flux corresponds to the number of nodes whose classification differs between two networks.

In our case, a Sankey diagram shows the classification of nodes into metastable states and transition region for each of the different recurrence networks computed from the tuning set $ \{ \varepsilon_{\nu} \} $.

\begin{figure}
\centering
\includegraphics[width = 1.15\textwidth, angle =90]{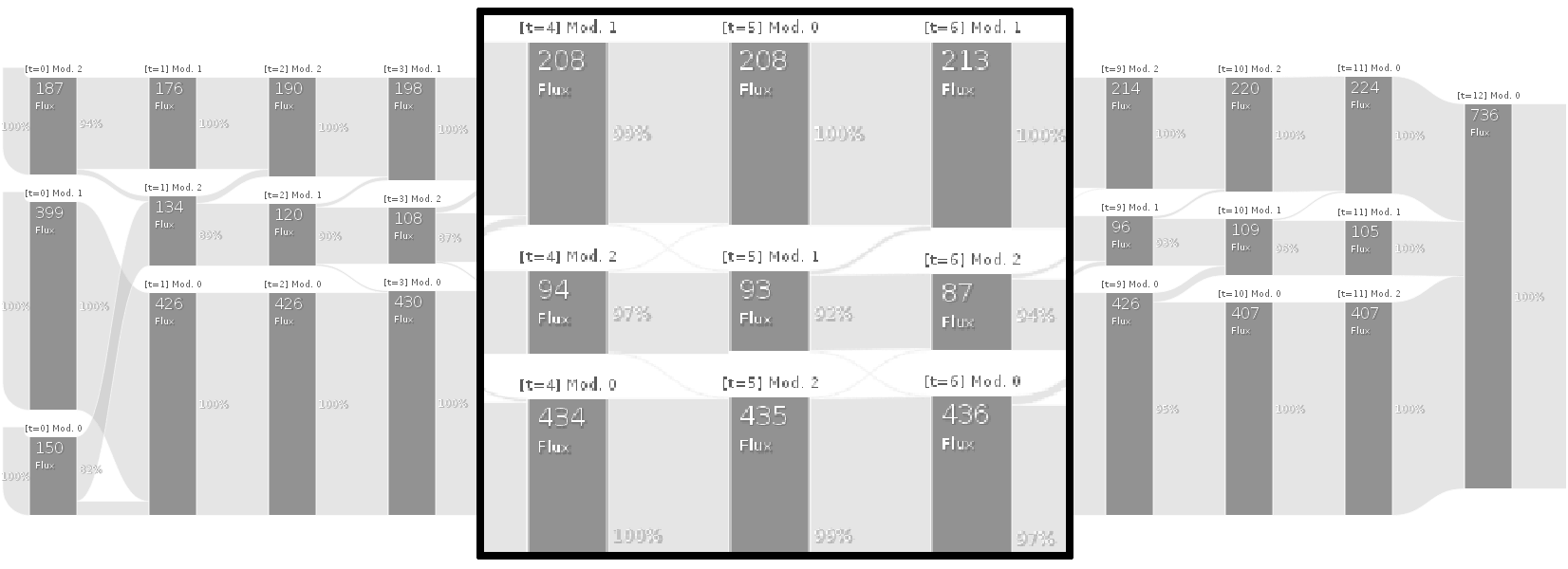}
\caption{\label{fig:dwpSankey}Sankey diagram showing the subgroup of recurrence networks (columns) with the same number of modules (expression \ref{eq:numberOfClustersTolerance}) and similar number of nodes (expression \ref{eq:sizeOfClustersTolerance}). Networks are computed from tuning set $ \{ \varepsilon_{\nu} \} $ (expression \ref{eq:sequenceRecurrenceThreshold}) on the state space constructed from a two well potential time series and embedding parameters $\tau = 7$ and $m = 2$. We suggest that this group of networks determines the recurrence threshold giving robust results about the dynamics of the time series analyzed.}
\end{figure}

Figure~\ref{fig:dwpSankey} shows the Sankey diagram used for the two well potential time series analysis of Section \ref{sec:Twp}. In this particular diagram, we zoom in the group of networks (columns) with the same number of modules (number of sections in every column), for which the number of nodes in each module (size of sections in every column) is almost the same (low flux of nodes from one column to another). Recurrence networks fulfilling conditions \ref{eq:numberOfClustersTolerance} and \ref{eq:sizeOfClustersTolerance} have a similar number and size of modules identified. These networks are used to set the final recurrence threshold used for the identification of metastable states in the two well potential time series. We suggest that these networks define a set of recurrence thresholds giving robust results about the dynamics of the time series analyzed.

\end{document}